\documentclass[12pt]{iopart}
\usepackage{iopams}
\usepackage{epsfig}
\begin{document}
%
%
\title[Two-particle interference in SQM and BQM]{Two-particle
interference in standard and Bohmian quantum mechanics}
\author{E Guay and L Marchildon}
\address{D\rm{\'{e}}partement de physique,
Universit\rm{\'{e}} du Qu\rm{\'{e}}bec,
Trois-Rivi\rm{\`{e}}res, Qc.\ Canada G9A~5H7}
\eads{\mailto{emilie\_guay@uqtr.ca}, \mailto{marchild@uqtr.ca}}
\begin{abstract}
The compatibility of standard and Bohmian quantum
mechanics has recently been challenged in the context
of two-particle interference, both from a theoretical and
an experimental point of view.  We analyze different
setups proposed and derive corresponding exact forms
for Bohmian equations of motion.  The equations are then solved
numerically, and shown to reproduce standard 
quantum-mechanical results.
\end{abstract}
\pacs{03.65}
\submitto{\JPA}
\maketitle
%
%
\section{Introduction}
Just about fifty years ago, Bohm proposed a new 
interpretation of quantum mechanics~\cite{Bohm}.  Although standard
quantum mechanics (SQM) is statistical and nondeterministic,
Bohmian quantum mechanics (BQM) is fully causal.
In BQM the wave-particle duality is resolved.
A quantum object is a particle with well-determined, although not accurately
known, position and momentum.  The quantum object's
wave characteristics are embodied in a quantum potential that
acts on the particle and is related to the wavefunction.  In a system
of $n$ particles with total wavefunction
$\Psi (\bi{r}_1, \ldots, \bi{r}_n, t)$, the velocity
of particle $i$ is taken to be~\cite{Bohm,Holland,BohmLivre}
\begin{equation}
\dot{\bi{r}}_{i} = \frac{\hbar}{m_{i}} 
\mbox{Im} \left[ \frac{\bi{\nabla}_{i}
\Psi (\bi{r}_1, \ldots, \bi{r}_n, t)}
{\Psi (\bi{r}_1, \ldots, \bi{r}_n, t)} \right] .
\label{vitesseBohm}
\end{equation}
In BQM, the probability density of finding particles $1, \ldots, n$
at points $\bi{r}_1, \ldots, \bi{r}_n$ is given, as in SQM,
by the absolute square of the total wavefunction.  The BQM probability
is subjective, that is, it expresses our ignorance of the particles'
true positions.  The continuity equation satisfied by the probability
density implies that the latter is consistent with Bohmian trajectories.
This means that if Bohmian particles are distributed according to the
absolute square of the wavefunction at time $t_0$, then their trajectories
will transform their probability distribution precisely as SQM predicts.
The probability of finding a given particle at a given position is 
the same in both interpretations.  Thus, statistical predictions of SQM 
and BQM should be indistinguishable.

Nevertheless, Ghose~\cite{Ghose1} as well as
Golshani and Akhavan~\cite{Gols1, Gols2, Gols3} recently
proposed experimental setups that allegedly lead to different
predictions for SQM and BQM.  An experiment along these lines has
been carried out by Brida \etal\cite{Brida},
and the results were interpreted as confirming SQM and
contradicting BQM\@.  All proposed experiments are based on 
two-particle interference, and make use of symmetrical
arrangements of either two or four slits.  The disagreement between
SQM and BQM is seen in the fact that SQM statistically allows 
some pairs of particles to reach detectors asymmetrically, while Bohmian
trajectories are claimed to forbid this.  These conclusions have been 
disputed by Struyve \etal\cite{Struyve} and by one of 
us~\cite{Marchildon1}.  The objection is that 
references~\cite{Ghose1,Gols1,Gols2,Gols3} make use of unwarranted
hypotheses on the initial distribution of the particles'
positions.

In this paper we first examine the double-slit and two-double-slit
setups proposed by Golshani and Akhavan, showing that in all 
important aspects the latter in fact reduces to the former.
Next we obtain the equations of motion exactly and point out general
properties of the velocities.  We then compute Bohmian trajectories 
numerically.  The trajectories are used to show explicitly how
the agreement between SQM and BQM comes about.
%
%
\section{Experimental setups}
\subsection{Double-slit setup}
Two-particle interference differs from one-particle interference
in that the interference pattern does not show up in the 
individual detection of a particle on a screen but in the 
joint detection of a pair of particles.  That is, the interference
pattern is a property of configuration space.

\begin{figure}[ht]
\begin{center}
\small
\begin{picture}(230,120)
\put(90,60){\vector(1,0){120}}
\put(90,0){\line(0,1){30}}
\put(90,40){\line(0,1){40}}
\put(90,90){\vector(0,1){30}}
\put(180,0){\line(0,1){120}}
\multiput(180,7.5)(0,10){12}{\line(2,-1){10}}
\put(80,60){\vector(0,1){25}}
\put(80,20){\vector(0,1){10}}
\put(80,50){\vector(0,-1){10}}
\put(0,60){\circle{15}}
\put(0,60){\makebox(0,0){S}}
\put(98,25){\makebox(0,0){B}}
\put(98,95){\makebox(0,0){A}}
\put(72,72.5){\makebox(0,0){$Y$}}
\put(70,35){\makebox(0,0){$2 \sigma_0$}}
\put(83,111){\makebox(0,0){$y$}}
\put(201,53){\makebox(0,0){$x$}}
\end{picture}
\end{center}
\caption{Double-slit setup.  Correlated pairs of
particles emitted from source S go through slits
A and B and are detected on a screen.}
\label{figure1}
\end{figure}
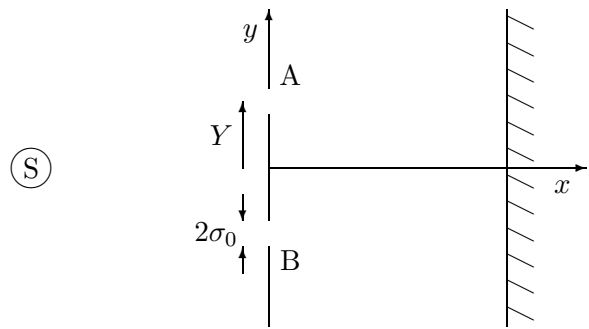

The double-slit setup proposed in \cite{Ghose1, Gols2} is 
shown in figure~\ref{figure1}. A pair of identical particles 
emitted by a source S impinges on a double-slit interferometer.
Just behind the slits the wavefunction is assumed to be well described
by plane waves with zero total momentum in the $y$-direction and identical
momenta in the $x$-direction.  Thus each particle goes through a 
different slit.  Since both particles are identical, the total
wavefunction is symmetrical (bosons) or antisymmetrical
(fermions) under particle permutation.  The edges of the slits
are assumed smooth enough to avoid treating diffraction of 
wave packets as these go through the
apertures~\cite[p.~177]{Holland}.  Each slit is taken
to generate a 
Gaussian wave form in the $y$-direction, while leaving 
the wave form unaffected in the $x$-direction.  Explicitly, let
particle $i$ emerge form slit A at $t=0$.  Its wavefunction 
is then written as
\begin{equation}
\fl \psi_{\rm{A}}(x_i, y_i, t=0) 
= \left( 2\pi \sigma_0^2 \right)^{-1/4}
\exp \left\{ - \frac{(y_i-Y)^{2}}{4\sigma_{0}^2} 
+ \rmi [k_x x_i + k_y (y_i-Y) ] \right\} .
\label{psiA0}
\end{equation}
We should point out that once we settle on equation~(\ref{psiA0})
as the one-particle wavefunction, no information on the particle's
state prior to $t=0$ is relevant that is not contained in~(\ref{psiA0}).
On the other hand, assuming free propagation with time we
get for $t>0$
\begin{eqnarray}
\nonumber \fl \psi_{\rm{A}}(x_i, y_i, t) 
= \left( 2\pi \sigma_t^2 \right)^{-1/4}
\exp \left\{ -(y_i -Y-\hbar k_y t/m)^{2}
/(4\sigma_{0}\sigma_{t}) \right\} \\
\times \exp \left\{ \rmi \left[k_x x_i 
+ k_y (y_i -Y-\hbar k_y t/(2m))- \hbar k_x^{2} t /(2m)
\right] \right\} , \label{psiA}
\end{eqnarray}
where
\begin{equation}
\sigma_{t} = \sigma_{0}\left(1 
+ \frac{\rmi \hbar t}{2 m \sigma_{0}^{2}}\right).
\label{sigmat}
\end{equation}

The two-particle total wavefunction is taken as
\begin{equation}
\Psi = N \left[ \psi_{\rm{A}}(x_{1}, y_{1}, t) 
\, \psi_{\rm{B}}( x_{2}, y_{2}, t) \pm
\psi_{\rm{A}}(x_{2}, y_{2}, t) 
\, \psi_{\rm{B}}(x_{1},  y_{1},  t) \right] , 
\label{psiTot2}
\end{equation}
where $N$ is a normalization constant and
\begin{equation}
\psi_{\rm{B}} (x, y, t) =\psi_{\rm{A}}(x, -y, t). \label{psiB}\\
\end{equation}
In (\ref{psiTot2}) the plus sign refers to bosons 
and the minus sign to fermions.
Note that the wavefunction $\Psi$ is symmetric 
(or antisymmetric) with respect to a
reflection in the $y=0$ plane.
\subsection{Two-double-slit setup}
\begin{figure}[ht]
\begin{center}
\small
\begin{picture}(190,120)(-92,0)
\put(20,60){\vector(1,0){150}}
\put(-20,60){\line(-1,0){150}}
\put(70,0){\line(0,1){30}}
\put(70,40){\line(0,1){40}}
\put(70,90){\line(0,1){30}}
\put(140,0){\line(0,1){120}}
\put(0,80){\vector(0,1){40}}
\multiput(140,7.5)(0,10){12}{\line(2,-1){10}}
\put(60,60){\vector(0,1){25}}
\put(0,60){\circle{15}}
\put(0,60){\makebox(0,0){S}}
\put(78,25){\makebox(0,0){B}}
\put(78,95){\makebox(0,0){A}}
\put(52,72.5){\makebox(0,0){$Y$}}
\put(7,111){\makebox(0,0){$y$}}
\put(161,53){\makebox(0,0){$x$}}
\put(-70,0){\line(0,1){30}}
\put(-70,40){\line(0,1){40}}
\put(-70,90){\line(0,1){30}}
\put(-140,0){\line(0,1){120}}
\multiput(-140,7.5)(0,10){12}{\line(-2,-1){10}}
\put(-60,70){\vector(0,1){10}}
\put(-60,100){\vector(0,-1){10}}
\put(0,15){\vector(1,0){70}}
\put(0,15){\vector(-1,0){70}}
\put(0,9){\makebox(0,0){$2 d$}}
\put(-78,25){\makebox(0,0){B$'$}}
\put(-78,95){\makebox(0,0){A$'$}}
\put(-50,85){\makebox(0,0){$2 \sigma_0$}}
\end{picture}
\end{center}
\caption{Two-double-slit setup.  Correlated pairs of
particles emitted from source S go through slits
A, B$'$ or B, A$'$ and are detected on screens.}
\label{figure2}
\end{figure}
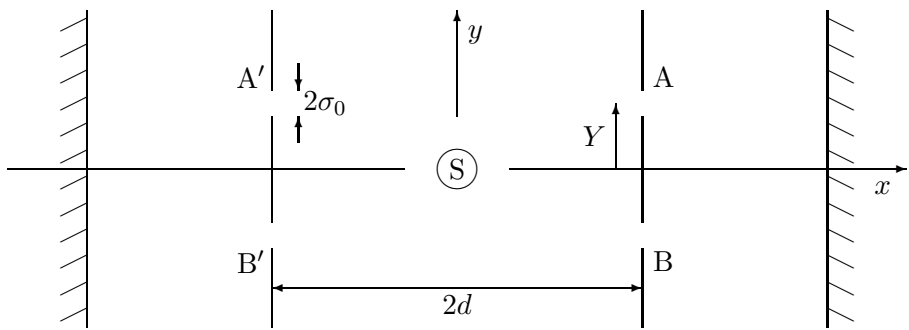

A two-double-slit setup was proposed in~\cite{Gols1}
and is shown in figure~\ref{figure2}.
A pair of correlated particles leaves source S
in a state of zero total momentum.  The particles
therefore either go through slits A and B$'$ or
through slits B and A$'$.  Accordingly, up to a
multiplicative factor the total
wavefunction was taken in~\cite{Gols1} as\footnote{This
wavefunction is symmetric with respect to a reflection
in the $y=0$ plane, for both bosons and fermions.
As with (\ref{psiTot2}), we could easily make it symmetric
for bosons and antisymmetric for fermions, by permuting
the sign factors of the third and fourth terms.  The
following argument would then be carried out just as
easily, except that the final correspondence between
equations~(\ref{4fentesA}) and~(\ref{4fentesB}) and
equation~(\ref{psiTot2}) would involve in the latter
both the plus and minus signs.}
\begin{eqnarray}
\nonumber \fl \Psi 
= \psi_{\rm{A}}(x_{1}, y_{1},  t) \,
\psi_{\rm{B}'} ( x_{2}, y_{2}, t) 
\pm \psi_{\rm{A}}(x_{2}, y_{2}, t) 
\, \psi_{\rm{B}'}(x_{1},  y_{1},  t) \\
\mbox{} + \psi_{\rm{A}'}(x_{2}, y_{2}, t) \,
\psi_{\rm{B}}(x_{1}, y_{1}, t)
\pm \psi_{\rm{A}'}(x_{1}, y_{1},  t) \, 
\psi_{\rm{B}}( x_{2}, y_{2}, t) , \label{psiTot4}
\end{eqnarray}
with
\begin{eqnarray}
\psi_{\rm{A'}}(x, y, t) = \psi_{\rm{A}}(-x, y, t) , \label{psiAA}\\
\psi_{\rm{B'}}(x, y, t) = \psi_{\rm{B}}(-x, y, t) 
= \psi_{\rm{A}}(-x, -y, t) .\label{psiBB}
\end{eqnarray}

%

It is not difficult to check that if the plus 
sign is picked in (\ref{psiTot4}), the
global wavefunction can be written as
\begin{equation}
\Psi = \cos[k_x (x_1 - x_2)] \Phi(y_1, y_2, t).
\end{equation}
If the minus sign is picked instead,
the cosine is simply replaced by a sine. 
Hence the $x_1$, $x_2$ dependence factors out 
and it is given by a real function. 
From equation~(\ref{vitesseBohm}), 
we immediately conclude that 
$\dot{x}_1 = 0 = \dot{x}_2$.  That is,
the Bohmian particles do not move in the 
$x$-direction.

The reason for this is that with the
one-particle wavefunctions written as 
in~(\ref{psiA}), (\ref{psiB}), (\ref{psiAA})
and~(\ref{psiBB}), equation~(\ref{psiTot4})
does not adequately represent the interference
situation depicted in figure~\ref{figure2}.
Indeed all one-particle wavefunctions
are infinitely extended in both the positive
and negative $x$-directions.  Clearly,
however, equations~(\ref{psiA}) and~(\ref{psiB})
for $\psi_{\rm{A}}$ and $\psi_{\rm{B}}$ are
intended to hold only where $x>d$, while
(\ref{psiAA}) and~(\ref{psiBB}) for 
$\psi_{\rm{A'}}$ and $\psi_{\rm{B'}}$ hold 
only where $x<-d$. The total wavefunction should
therefore be written as
\begin{equation}
\Psi = N' [\psi_{\rm{A}}(x_{1}, y_{1},  t) \, 
\psi_{\rm{B'}}( x_{2}, y_{2}, t) +
\psi_{\rm{A'}}(x_{2}, y_{2}, t) \,
\psi_{\rm{B}}(x_{1}, y_{1}, t)] \label{4fentesA}
\end{equation}
if $x_1 > d$ and $x_2 < -d$, and as
\begin{equation}
\Psi = \pm N' [ \psi_{\rm{A}}(x_{2}, y_{2}, t) \,
\psi_{\rm{B'}}(x_{1}, y_{1}, t) +
\psi_{\rm{A'}}(x_{1}, y_{1}, t) \, 
\psi_{\rm{B}}(x_{2}, y_{2}, t)] 
\label{4fentesB}
\end{equation}
if $x_1 < -d$ and $x_2 > d$. 
The two expressions of the total wavefunction
do not overlap in configuration space, whence
Bohmian trajectories associated with one are
completely independent of the other.  
As it should, (\ref{4fentesA})
and~(\ref{4fentesB}) transform into each other
(with a plus sign for bosons and a minus sign
for fermions) under particle permutation.

It is easy to check that, up to a constant
multiplicative factor, equation~(\ref{4fentesA})
can be obtained from~(\ref{psiTot2}) 
(with the plus sign) through
the substitution $x_2 \rightarrow -x_2$.
From~(\ref{vitesseBohm}) we see at once that
all components of velocities are the same with
both wavefunctions, except that $\dot{x}_2$ is 
transformed into $- \dot{x}_2$.  We conclude that
Bohmian trajectories in the two situations are
in one-to-one correspondence, with $x_2$ being
reflected in the $yz$-plane.

Exactly the same argument shows that Bohmian
trajectories computed with~(\ref{4fentesB})
and~(\ref{psiTot2}) are also in one-to-one 
correspondence, with $x_1$ now being
reflected in the $yz$-plane.
\section{SQM and BQM predictions}
In the remainder of this paper we shall
discuss the double-slit setup only, since
results pertaining to the two-double-slit
setup can be obtained by straightforward
transformation.

Let one pair of particles leave the slits
at $t=0$ and arrive at detectors at time $t$
(both particles have $x$-momentum equal to
$\hbar k_x$).  In SQM, the probability of finding the
particles on detectors at points $y_1$ and $y_2$
is given by:
\begin{equation}
P(y_1, y_2, t) = | \Psi(y_1, y_2, t)|^{2} .
\end{equation}

Suppose first that $t$ is such that
$|\sigma_t| \approx \sigma_0$.  Then the
spreading of the wave packets is not very
important. Therefore, the probability of 
finding both particles on the same side of
the $x$-axis is very low.  The particles will be
detected on both sides of the $x$-axis, rather
symmetrically if $\sigma_0$ is much smaller than $Y$ 
(see figure~\ref{figure1}).  Moreover, as one-particle
wavefunctions overlap very little, interference effects
will be negligible. 

Assume now that $|\sigma_t| \gg \sigma_0$.
This can be obtained either by taking detectors
further to the right, or by reducing the value
of $k_x$.  Then wavefunction overlap becomes
important and interference effects begin to show up.
Detection becomes more and more asymmetrical,
and both members of a pair can even be detected on 
the same side of the $x$-axis.


In BQM, particle velocities can be computed 
from equation~(\ref{vitesseBohm}) and 
wavefunction~(\ref{psiTot2}).  The calculation
somewhat simplifies if $k_y = 0$, which we
henceforth assume.  Substituting~(\ref{psiA})
and~(\ref{psiB}) into~(\ref{psiTot2}) and
discarding multiplicative factors that do not
depend on $\bi{r}_1$ and $\bi{r}_2$, we find that
\begin{eqnarray}
\fl \Psi \sim \exp \left\{ \rmi \left[ k_x
(x_1 + x_2) \right] \right\}
\left\{ \exp \left[ - \frac{y_1^2 + y_2^2
+ 2 Y (y_2 - y_1)}{4 \sigma_0 \sigma_t} 
\right] \right. \\
\pm \left. \exp \left[ - \frac{y_1^2 + y_2^2
- 2 Y (y_2 - y_1)}{4 \sigma_0 \sigma_t} 
\right] \right\} .
\end{eqnarray}
Making use of~(\ref{vitesseBohm}) we immediately
see that
\begin{equation}
\dot{x}_{1} =  \frac{\hbar k_x}{m} = \dot{x}_{2}. \label{vitesses_X}
\end{equation}

The $y$-components are trickier.  Discarding
now the $x_1$ and $x_2$ dependence of the wave
function and making use of~(\ref{sigmat}), 
we can write
\begin{equation}
\Psi \sim \exp (-f) \left\{ \exp(-g)
\pm \exp(g) \right\} ,
\end{equation}
where
\begin{eqnarray}
f = \left\{ 4 \sigma_0^2 \left[ 1
+ \left( \frac{\hbar t}{2 m \sigma_0^2} \right)^2
\right] \right\} ^{-1} \left[ 1
- \frac{\rmi \hbar t}{2 m \sigma_0^2} \right]
\left( y_1^2 + y_2^2 \right) , \\
g = \left\{ 4 \sigma_0^2 \left[ 1
+ \left( \frac{\hbar t}{2 m \sigma_0^2} \right)^2
\right] \right\} ^{-1} \left[ 1
- \frac{\rmi \hbar t}{2 m \sigma_0^2} \right]
2 Y ( y_2 - y_1 ) .
\end{eqnarray}
From~(\ref{vitesseBohm}) we get
\begin{equation}
\dot{y}_1 = \frac{\hbar}{m} \, \mbox{Im}
\left\{ - \frac{\partial f}{\partial y_1}
+ \left( \frac{\partial g}{\partial y_1} \right)
\frac{ - \exp(-g) \pm \exp(g)}
{ \exp(-g) \pm \exp(g)} \right\} .
\end{equation}
Straightforward manipulations and use of trigonometric
identities finally yield
\begin{equation}
\dot{y}_{1}= - \frac{ 2 \hbar Y m \sigma_0^2 
\sin( \hbar t \alpha )\pm Y \hbar^2 t
\sinh(2 m \sigma_0^2 \alpha ) }
{\left(\hbar^2 t^2+4 m^2 \sigma_0^4\right) 
[\cos( \hbar t \alpha) \pm 
\cosh (2 m \sigma_0^2 \alpha )]} 
+ \frac{\hbar^2 t {y_1}}{\hbar^2 t^2 
+ 4 m^2 \sigma_0^4} ,
\label{vitesse_Y_1}
\end{equation}
where the upper sign is for bosons, the
lower sign for fermions and
\begin{equation}
\alpha = \frac{2 Y m \left({y_1}-{y_2}\right)}
{\hbar^2 t^2+4 m^2 \sigma_0^4} .
\end{equation}
Similarly,
\begin{equation}
\dot{y}_{2}=\frac{ 2 \hbar Y m \sigma_0^2 
\sin( \hbar t \alpha )\pm Y \hbar^2 t
\sinh(2 m \sigma_0^2 \alpha ) }
{\left(\hbar^2 t^2+4 m^2 \sigma_0^4\right) 
[\cos( \hbar t \alpha)\pm 
\cosh (2 m \sigma_0^2 \alpha )]} 
+ \frac{\hbar^2 t {y_2}}{\hbar^2 t^2 
+ 4 m^2 \sigma_0^4} . \label{vitesse_Y_2}
\end{equation}

When $t$ is large enough, $\dot{y}_1$ 
and $\dot{y}_2$ are dominated by the
last term in~(\ref{vitesse_Y_1})
and~(\ref{vitesse_Y_2}).  The behaviour of
fermions and bosons therefore coincide.

It is easy to check that velocity components
satisfy the following relations:
\begin{eqnarray}
\dot{y}_i(x_1, y_1; x_2, y_2,; t) 
= -\dot{y}_i(x_1, -y_1; x_2, -y_2; t) 
\qquad (i = 1, 2) , \label{egalVitesse1}\\
\dot{y}_1(x_1, y_1; x_2, y_2,; t) 
= -\dot{y}_2(x_1, -y_2; x_2, -y_1; t).
\label{egalVitesse3}
\end{eqnarray}
Furthermore, the $y$-component of the 
centre-of-mass coordinate satisfies
\begin{equation}
\dot{y}(t) \equiv \frac{\dot{y}_{1}(t)+ \dot{y}_{2}(t)}{2} =\frac{\hbar^2 t
{y}}{\hbar^2 t^2+4 m^2 \sigma_0^4} .
\end{equation}
This is readily integrated as~\cite{Gols2}
\begin{equation}
y(t) = y(0) \left\{ 1 + \left( \frac{\hbar t}
{2m \sigma_0^2} \right)^2 \right\} ^{1/2}
= y(0) \frac{|\sigma_t|}{\sigma_0} . \label{Vcentre}
\end{equation}

If two particles emerge from the slits at heights symmetrical 
with respect to the $x$-axis, then $y(0) = 0$.  
Equation~(\ref{Vcentre}) implies that $y(t) = 0$ for 
all $t$, and the particles will necessarily be
detected symmetrically.  Likewise if $y(0)$ is very small,
specifically, much smaller than $\sigma_0$, $y(t)$ should
remain small enough so that the two particles will
be detected almost symmetrically.

From such observations, Ghose~\cite{Ghose1} and Golshani and 
Akhavan~\cite{Gols3} argued that BQM could make experimental
predictions beyond what SQM allows.  One would only have
to prepare a number of pairs each with $|y(0)| \ll \sigma_0$,
and detect them at time $t$.  We have seen that, if
$|\sigma_t| \gg \sigma_0$, SQM predicts highly asymmetrical
detection.  Yet with $y(0)$ small enough, BQM would predict
highly symmetrical detection.

The flaw in the argument was pointed out in~\cite{Struyve,
Marchildon1}, where it was shown that such a selection of
$y(0)$ values is incompatible with Bohm's assumptions.
It is instructive to make the argument fully quantitative.
In Bohm's theory, the probability distribution of
particle positions is given by the absolute square of
the wavefunction.  Making use of~(\ref{psiTot2}),
(\ref{psiA0}) and~(\ref{psiB}), we see that at $t=0$
the distribution of $y$-coordinates is given by ($k_y = 0$)
\begin{equation}
P(y_1, y_2, 0) = |N|^2 (2 \pi \sigma_0^2)^{-1}
\left\{ F + G \pm 2H \right\} ,
\end{equation}
where
\begin{eqnarray}
F = \exp \left\{ - \frac{(y_1 - Y)^2 + (y_2 + Y)^2}
{2 \sigma_0^2} \right\} , \\
G = \exp \left\{ - \frac{(y_2 - Y)^2 + (y_1 + Y)^2}
{2 \sigma_0^2} \right\} , \\
H = \exp \left\{ - \frac{y_1^2 + y_2^2 + 2 Y^2}
{2 \sigma_0^2} \right\} .
\end{eqnarray}
Straightforward integration shows that the probability is
normalized to one if
\begin{equation}
|N|^2 = \frac{1}{2} \left\{ 1 \pm \exp \left( -
\frac{Y^2}{\sigma_0^2} \right) \right\}^{-1} .
\end{equation}

It is easy to check that $\langle y_1 + y_2 \rangle = 0$.
The standard deviation of $y(0)$ then follows from an evaluation
of $\langle (y_1 + y_2)^2 \rangle$.  All integrations are
elementary, and we obtain
\begin{equation}
\Delta y(0) = \frac{1}{2} \sqrt{\langle (y_1 + y_2)^2 \rangle}
= \frac{\sigma_0}{\sqrt{2}} . \label{ecart-type}
\end{equation}
Hence it is not possible, in Bohm's theory, to select
initial positions so that $|y(0)| \ll \sigma_0$.
Any such selection scheme amounts either to pick a different
initial wavefunction, or to make assumptions on the
distribution of true particle positions different from
Bohm's original ones.
\section{Trajectories}
To investigate Bohmian trajectories of pairs of particles
governed by wavefunction~(\ref{psiTot2}), we have computed
them numerically.  A fourth- and fifth-order Runge-Kutta
algorithm was implemented to solve differential 
equations~(\ref{vitesse_Y_1}) and~(\ref{vitesse_Y_2}),
both in Mathematica and through a special purpose program we
wrote in C.  Results displayed all use $m$ equal to the mass
of the electron, $\sigma_0 = 10^{-6}\, \mbox{m}$, 
$Y = 5 \, \sigma_0$ and $k_y = 0$.  The distance between
slits and detectors is set at $0.2 \, \mbox{m}$.
From~(\ref{vitesses_X}) we see that the time needed for
a particle to reach the detector is inversely proportional
to $k_x$.

\begin{figure}[h]
\begin{center}
\epsfbox{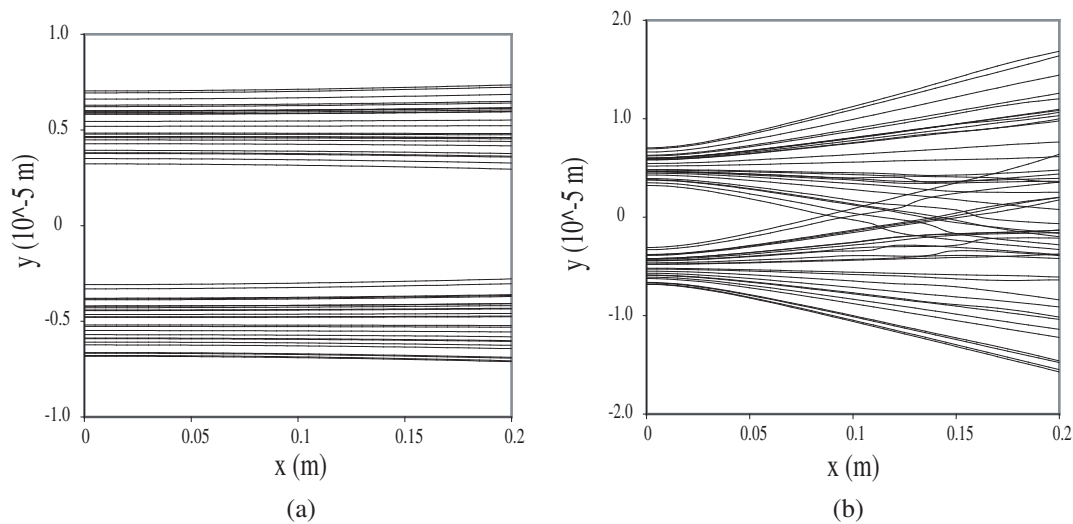}
\end{center}
\caption{\label{graph1}25 pairs of trajectories for (a) 
$\hbar k_x /m = 2\times10^7 \, \mbox{m/s}$ and (b)
$\hbar k_x /m = 2\times10^6 \, \mbox{m/s}$.}
\end{figure}

Figure~\ref{graph1} shows a number of pair trajectories
for two different values of $k_x$, with the result that 
$|\sigma_t| = 1.16 \, \sigma_0$ in (a) and $|\sigma_t| =
5.88 \, \sigma_0$ in (b).  For each pair, $y$-coordinates
were picked randomly according to Gaussian distributions
with standard deviation $\sigma_0$, one centred about $Y$
and the other centred about $-Y$.  In (a) the Bohmian 
trajectories are almost horizontal straight lines, which
bears out the fact that wave packets have spread very
little.  In (b), on the other hand, Bohmian trajectories
spread and a number of pairs display considerable
interference.  Some particles coming out of the upper slit end
up below the $x$-axis, and some coming out of the lower
slit end up above.

\begin{figure}[h]
\begin{center}
\epsfbox{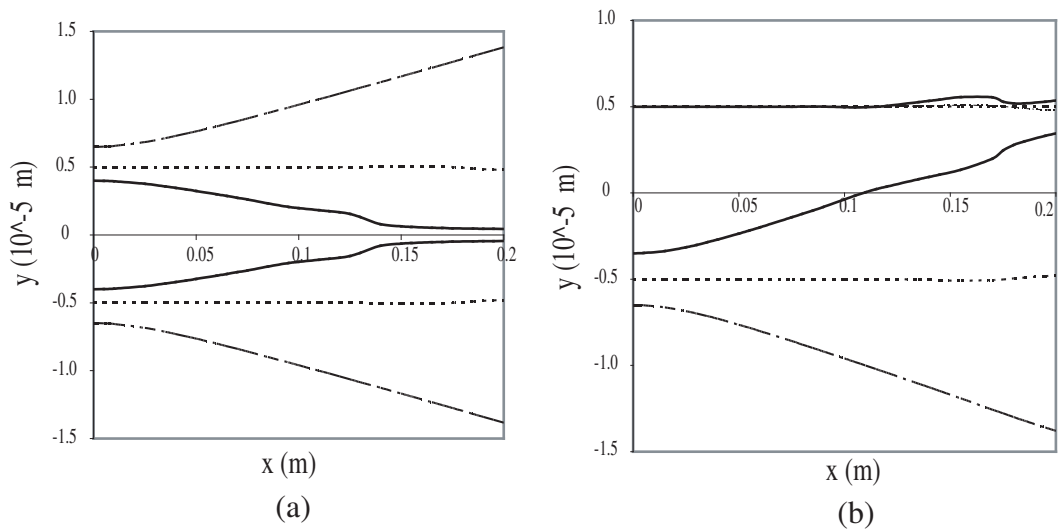}
\end{center}
\caption{\label{graph2} Three pairs of trajectories with
(a) symmetrical and (b) asymmetrical initial conditions.}
\end{figure}

Figure~\ref{graph2} shows a much smaller number of pairs,
this time labelled independently.  In (a) the initial $y$-coordinates
of both particles are picked symmetrically, that is,
$y(0) = 0$.  Trajectories remain symmetrical,
as equation~(\ref{Vcentre}) predicts.  In (b), on the other hand,
the initial $y$-coordinates of the upper particles are set
at $y=Y$, while the $y$-coordinates of the lower particles
are picked as $y = -Y + 1.5 \, \sigma_0$ (solid line), 
$y = -Y$ (dotted line) and $y = -Y - 1.5 \, \sigma_0$ 
(dotted-dashed line), respectively.  The particle leaving
from $y = -Y + 1.5 \, \sigma_0$ clearly
ends up above the $x$-axis, and it gets
arbitrarily far from the axis if allowed to go on.

To sum up, the statistical distribution of
Bohmian trajectories is fully consistent with the
predictions of standard quantum mechanics.  Moreover,
pairs of Bohmian particles can be detected on the
same side of the setup's symmetry plane, and
arbitrarily far from the plane.

In the experiment reported in reference~\cite{Brida},
pairs of photons generated by parametric 
down-conversion were allowed to go through a 
double-slit setup of the type shown in 
figure~\ref{figure1}.  A number of pairs were detected
on the same side of the symmetry plane.  On the basis
of symmetric photon trajectories obtained
in~\cite{Ghose2}, and following the analysis of
electronlike trajectories carried out 
in~\cite{Ghose1, Gols3}, this result was interpreted as 
confirming SQM against BQM.

It is clear that our analysis leads to a
different picture.  We have shown that nothing prevents
Bohmian particles from being detected on the same side
of the symmetry plane.  Hence, under the assumption 
that electronlike trajectories are in that respect
relevant to photons, the results of~\cite{Brida} should
be interpreted as confirming both SQM and BQM.
\section{Conclusion}
In this paper, double-slit and two-double-slit
setups for two-particle interference were analyzed
and shown to be essentially equivalent.
Bohmian equations of motion for pairs of bosons
or fermions were then obtained and numerically
integrated.  The statistical distribution of Bohmian
trajectories turns out to reproduce standard
quantum-mechanical results.  Members of a given pair 
are not restricted to remain on different sides of
a symmetry plane.  Relevant experimental results are 
therefore consistent with both SQM and BQM.

\ack{We thank P~Ghose and M~Genovese for comments.
One of us (EG) is grateful to the Natural Sciences
and Engineering Research Council of Canada for the award of a 
postgraduate scholarship.}
\end{document}